ABOUT THE MECHANISM OF INTERFERENCE OF SILVER STAINING
WITH PEPTIDE MASS SPECTROMETRY

Sophie Richert 1, Sylvie Luche 2, Mireille Chevallet 2, Alain Van Dorsselaer 1,
Emmanuelle Leize-Wagner 1 and Thierry Rabilloud 2*

1: Laboratoire de Spectrométrie de Masse Bio-Organique, UMR CNRS 7509, ECPM, 25 rue
Becquerel, 67008 Strasbourg Cedex

2:  CEA- Laboratoire de Bioénergétique Cellulaire et Pathologique, EA 2943,  DRDC/BECP
CEA-Grenoble, 17 rue des martyrs,  F-38054 GRENOBLE CEDEX 9,  France

(Running title): silver staining interference with mass spectrometry

*: to whom correspondence should be addressed

Correspondence :
Thierry Rabilloud, DRDC/BECP
CEA-Grenoble, 17 rue des martyrs,
F-38054 GRENOBLE CEDEX 9
Tel (33)-438-78-32-12
Fax (33)-438-78-51-87
e-mail: Thierry.Rabilloud@cea.fr

Abbreviations: ESI: electrospray ionisation; MALDI: Matrix-Assisted Laser Desorption and
Ionization; MS:  mass spectrometry; RuBPS: ruthenium II tris (bathophenanthroline
disulfonate).

Abstract

The mechanism by which silver staining of proteins in polyacrylamide gels interferes with mass spectrometry of peptides produced by proteolysis has been investigated. It was demonstrated that this interference increases with time between silver staining and gel processing, although the silver image is constant. This suggested an important role of the formaldehyde used in silver staining development in this interference process. Consequently, a formaldehyde-free staining protocol has been devised, using carbohydrazide as the developing agent. This protocol showed much increased peptide coverage and retained the sensitivity of silver staining. These results were however obtained at the expense of an increased background in the stained gels and of a reduced staining homogeneity.

1 Introduction

Proteomics analyses impart special constraints on the detection techniques used after two-dimensional gel electrophoresis. For example, in addition to the standard sensitivity issues, proteomics put special emphasis on the interference with the microcharacterization techniques used afterwards, i.e. in most cases mass spectrometry analysis by MALDI-MS or nanoESI-MS/MS. For these reasons, silver staining, which is still the most sensitive non-radioactive detection technique, is not optimal, because losses in peptide masses or problems in nanoESI-MS/MS have been documented [1, 2].
Apart from silver staining, the most widely used detection technique relies on colloidal Coomassie Blue [3]. While this technique provides much better results in terms of linearity, homogeneity and interference with mass spectrometry, its sensitivity is much too low. Thus, either high loads of proteins must be used, at the risk of protein losses by precipitation, or the analysis is restricted to major proteins only.

Heavy salt precipitation has also been used as a mild and sensitive detection technique. The most sensitive technique in this family is represented by the zinc-imidazole protocol [4]. This technique performs extremely well for mass spectrometry, as there is no fixation of the proteins and therefore maximal accessibility of the cleavage sites to digestion and maximal extractability of the resulting peptides. Despite these very strong points, this technique suffers from some drawbacks that have limited its widespread use. Among those are its poor ability to detect some classes of proteins, such as low molecular weight proteins and glycoproteins [5] and its poor contrast that makes precise band or spot excision difficult.

Another detection technique which appeared recently is based on fluorescence, either of purely organic probes (e.g. Sypro Orange and Sypro Red) or of ruthenium or europium complexes (Sypro Ruby, Sypro Rose and RUBPS complex). Organic probes show a sensitivity slightly better than colloidal Coomassie blue, while ruthenium complexes have a sensitivity slightly inferior to that of silver staining but a much improved compatibility with mass spectrometry [6, 7]. These methods, however, require expensive hardware to detect the fluorescent signal. Moreover, the uniform illumination required for spot excision leads to some fading of fluorescence and thus losses of performance. This effect is much more pronounced for purely organic fluorophores than for metal-organic fluorophores. In addition, UV tables used sometimes for this illumination are rather hazardous. Last but not least, many fluorescent detection protocols are not steady-state, so that the stained gel cannot be kept for future use if convenient.

Thus, the best situation among the methods available to date would be to combine the sensitivity of silver staining, its simplicity of visualization for spot excision and its stability over time, but to improve its compatibility with peptide mass analysis. Interesting methods in this trend have been devised by destaining protocols performed after staining, either with ferricyanide [8] or with hydrogen peroxide [9]. While these methods allow an important improvement in peptide recovery over undestained gels, the peptide coverage obtained are still inferior to those obtained with Coomassie Blue, zinc imidazole or fluorescent probes. We therefore decided to investigate the chemical mechanism of the interference of silver staining with peptide analysis with mass spectrometry techniques, in order to devise improved silver staining or post-staining processing protocols.

2. Materials and methods

2.1. Gel electrophoresis
Proteins were separated by SDS-PAGE, either in the standard Tris-glycine system, or in the Tris-taurine [10]. Molecular weight markers (Bio Rad, Broad range) were diluted 200 and 2000 fold in SDS sample buffer to reach a concentration range of 10 and 1 ng/µl for each band. the required volumes were loaded on top of a 10% gel to give the adequate concentrations, ranging from 5 to 400 ng per band.

2.3. Detection of proteins after electrophoresis
Detection of proteins by colloidal Coomassie blue or fluorescent complexes was carried out according to published protocols [3, 6]. For silver staining, fast and long methods with silver nitrate were used [11, 12]. For comparison, ammoniacal silver methods were also used [13]. The formaldehyde-free staining protocol was based on the fast silver staining methods, but used carbohydrazide instead of formaldehyde as the developing agent. Other chemicals, i.e. ortho- and para-phenylenediamine, phenidone, hydroxylamine, semithiocarbazide and semicarbazide were also tested. The silver staining protocol is as follows:
1. Fix the gels ( 1 h + overnight) in 5% acetic acid/30% ethanol (v/v). If a shorter time is preferred, gels can be fixed in 10% acetic acid/30% ethanol (v/v) for 3 x 30 min.
2. Rinse in water for 4 x 10 min.
3. To sensitize, soak gels for 1 min (1 gel at a time) in 0.8 mM sodium thiosulfate
4. Rinse 2x 1 min in water .
5. Impregnate for 30-60 min in 12 mM silver nitrate (0.2g/l). The gels may become yellowish at this stage.
6. Rinse in water for 5-15 s

7. Develop image (1-2 min) in 3% potassium carbonate containing 300-500 µM carbohydrazide and 125 µl 10% sodium thiosulfate per liter.
8. Stop development (30-60 min) in a solution containing 40 g of Tris and 20 ml of acetic acid per liter.
9. Rinse with water (several changes) prior to drying or densitometry.

Spot destaining was performed according to previously-described methods 8,9 either immediately after silver staining (including the stop and wash steps) or 2 days after silver staining. In control experiments, 3mM formaldehyde was used in place of carbohydrazide for image development.

2.4. Mass spectrometry analysis
Stained proteins spots or bands were excised (on a UV table for fluorescent detection), and shrunk in 1 ml of 50% ethanol for 2 hours.

In gel digestion :
Each gel slice was cut into small pieces with a scalpel, washed with 100 µl of 25 mM NH4HCO3 and dehydrated with 100 µl of acetonitrile. This operation was repeated twice. Reduction was achieved by 1 hour treatment with 10mM DTT at 57°C. Alkylation reaction was performed by 25mM Iodoacetamide for 45 min at room temperature, protected from light. Finally, gel spots were washed 3 times for 5 minutes again alternately with 25mM ammonium carbonate and acetonitrile. Gel pieces were completely dried with a Speed Vac before tryptic digestion. The dried gel volume was evaluated and three volumes of trypsin (Promega, V5111), 12.5 ng/µl, in 25 mM NH4HCO3 (freshly diluted) were added. The digestion was performed at 35°C overnight with 5 to 10 µl of buffer. The gel pieces were centrifuged and 5 µl of 25% H2O/70% Acetonitrile/5% HCOOH were added to extract peptides. The mixture was sonicated for 5 min. and centrifuged. The supernatant was recovered and the operation was repeated once. For MALDI-MS analysis, the supernatant volume was reduced under nitrogen flow to 4 µl, 1 µl of H2O/5% HCOOH were added and 0.5 µl of the mix were used for the analysis. For nano LC-MS/MS, the supernatant solvent was completely evaporated in order to remove all acetonitrile from the sample. Then, 10 µl of H2O/5% HCOOH were added and injected in the nano HPLC system.

MALDI-MS : For MALDI mass spectrometry, mass measurements were carried out on a Bruker BIFLEX tion was dried under vacuum. The sample was washed one to three times by applying 1µl of aqueous HCOOH (5 %) solution on the target and then flushed after a few seconds. In positive mode, internal calibration is performed with tryptic peptides coming from autodigestion of trypsin, with respectively monoisotopic masses at m/z = 842.51; m/z =

1045.564; m/z = 2211.105. Monoisotopic peptide masses were assigned and used for databases searches.

These files were then fed into the search engine MASCOT (Matrix Science, London, UK). The data were searched against NCBI non-redundant protein sequence database with trypsin plus potentially two missed cleavages. All proteins present in NCBI were taken into account without any pI and MW restrictions. Some variable modifications are taken into account, like methionine oxidation, cysteine carbamidomethylation. The peptide mass error was limited to 50 ppm.

NanoLC-MS/MS : Nanoscale capillary liquid chromatography-tandem mass spectrometric (LC-MS-MS) analysis of the digested proteins were performed using a CapLC capillary LC system (Micromass, Manchester, UK) coupled to a hybrid quadrupole orthogonal acceleration time-of-flight tandem mass spectrometer (Q-TOF II, Micromass). The LC-MS union was made with a PicoTip (New Objective, Woburn,.MA) fitted on a ZSPRAY (Micromass) interface. Chromatographic separations were conducted on a reversed-phase (RP) capillary column (Pepmap C18, 75µm i.d., 15 cm lenght, LC Packings) with a 200 nL/min flow. The gradient profile used consisted of a linear gradient from 95% A (H2O / 0.05% HCOOH) to 45% B ( acetonitrile / 0.05% HCOOH) in 35 min. followed by a linear gradient to 95% B in 1 min. Mass data acquisitions were piloted by MassLynx software (Micromass, Manchester, UK) using automatic switching between MS and MS/MS modes. The internal parameters of Q-TOF II were set as follows. The electrospray capillary voltage was set to 3.0 kV, the cone voltage set to 30 V, and the source temperature set to 80°C. The MS survey scan was m/z 300-1500 with a scan time of 1 s and a interscan time of 0.1s. When the intensity of a peak rose above a threshold of 8 counts, tandem mass spectra were acquired. Normalized collision energies for peptide fragmentation was set using the charge-state recognition files for +1, +2 and +3 peptides ions. The scan range for MS/MS acquisition was from m/z 50 to 1500 with a scan time of 1 s and a interscan time of 0.1s. Fragmentation was performed using argon as the collision gas and with a collision energy profile optimized for various mass ranges of precursor ions.

Mass data collected during a nanoLC-MS/MS analysis were processed and converted into a .PKL file to be submitted to the search software MASCOT (Matrix Science, London, UK). Searches were done with a tolerance on mass measurement of 0.25 Da in MS mode and 0.5 Da in MS/MS mode.

3. Results

3.1. Sensitivity evaluation

This test was carried out by staining serial dilutions of protein markers separated by SDS PAGE, as shown in Figure 1. The formaldehyde-free protocol resulted in extremely fast development (1-2 minutes), but also gave a dark background that can be partly filtered at the scanning step. In addition, the staining was located at the very surface of the gel and was almost undetectable by scanning the gel in its thickness, while the formaldehyde silver staining gave a much thicker silver deposit (data not shown). The sensitivity was roughly equivalent for the two staining methods, and 10 ng were detectable with both methods. Rather similar results were obtained with semicarbazide and semithiocarbazide. However, these chemicals were less easy to use and gave more erratic results. We could not obtain any adequate staining with phenidone, hydroxylamine or phenylene diamine.

3.2. Mass spectrometry

The peptide mass fingerprinting by MALDI-MS and nanoLC ESI-MS/MS experiments were carried out at least in triplicate to get an average value of sequence coverage. Typical results are shown in figures 2 and 3 and in Table 1. From these data, several important trends emerged:

The classical silver staining with formaldehyde in the developer gives much lower sequence coverage, even after destaining, than the more benign but less sensitive methods (colloidal coomassie blue, fluorescent ruthenium complexes)
The silver staining protocol with carbohydrazide gives much improved sequence coverage over the classical formaldehyde silver stain.
The sequence coverage is increased if the stained bands are destained by the ferricyanide method [8] . The resulting coverage is then very close to the one obtained for example with Coomassie Blue.

It must be mentioned that the nanoLC ESI-MS/MS experiments were carried out on both an ion trap (Esquire; Bruker) and a Q-TOF II instrument (Micromass). Comparable results were obtained in both instruments, thereby showing the versatility of the methods described here.

In order to further investigate the mechanisms at play in the silver staining interference process, additional experiments were carried out by varying the destaining protocol. Briefly, the gels were either destained the same day as staining or left for another 48 hours in a water bath before destaining. While the sequence coverage was hardly effected by this additional period of time between staining and destaining in the case of the carbohydrazide-silver stain, the coverage decreased dramatically when the gels were left for 48 hours in water after formaldehyde-silver staining and before band excision and destaining. As a matter of fact, no protein could be identified from a 200 ng band after this 48 hours bath, while this amount of protein gave quite adequate identification when destaining is carried out on the same day as staining (see Table 1). In another experiment, the silver metal-containing surface layer was scraped from the gel plugs, which were then used without destaining. The sequence coverage was slightly better than the one observed on undestained gels, but clearly inferior to the one reached with additional destaining. Destaining was therefore adopted as a routine procedure prior to MS analysis.

To determine the useful threshold for the different detection methods, sensitivity experiments were carried out on serial dilutions of proteins separated on gels and then detected by different techniques. The results are shown on Table 2. They clearly demonstrate that the ferricyanide destaining protocol is clearly superior to the one based on hydrogen peroxide. More importantly, these data also show that the useful sensitivity is close to 0.5 picomoles, although much lower levels of proteins can be detected. This limit does not seem to be related to the silver staining per se, as the same limit is also encountered for the ruthenium-stained gels. It must be reminded, however, that this limit is obtained in SDS PAGE with rather wide lanes (ca 8 mm). Due to the focusing effect that allows a much smaller gel plug to be analyzed, the limits seems to be lower for two-dimensional gels, close to 100-200 fmol.

Finally, to evaluate the usefulness of the carbohydrazide method in a real proteomics experiment, this method was tested on a complex cell extract separated by 2D gel electrophoresis. The results are shown on figure 4. Apart from the high background, which has been already mentionned and can be corrected, a second drawback appears, which is a limited staining homogeneity, in the sense that some proteins which are easily detected with a classical silver staining are hardly detected with the carbohydrazide protocol (compare panel 4A vs. 4B). However, the carbohydrazide staining method is still fairly more sensitive than colloidal Coomassie Blue (compare panel 4B vs. 4C) , and delivered results in less than 24 hours, instead of the 48 hours requested for optimal colloidal Commassie Blue staining

4. Discussion

The results described above provided some precise insights about the chemical mechanisms involved in the interference phenomenon induced by silver staining. While the exact role of the silver metal image itself is not very clear yet, previous blotting experiments from undestained gels [14], in which whole proteins are able to cross the silver image, strongly suggest that the metal image itself is not the major interfering entity. The difference between the classical and formaldehyde-free protocols strongly suggests that formaldehyde is a major interfering compound. The mechanism of this interference is probably some crosslinking between aminoacids reactive side chains (lysine, cysteine, and to a lesser extent serine and threonine), leading to protein reticulation by methylene bridges, in a way quite similar to that observed during histochemical fixation [15]. It is very likely that this crosslinking occurs in a two-step mechanism. The first step involves grafting of a hydroxymethyl group on a reactive side chain, and is followed by a second substitution leading to crosslinking. This two-step mechanism would explain the slow worsening in interference induced upon simple storage in water.

However, the improvement shown by destaining in formaldehyde-free, silver stained gels also shows that formaldehyde is not responsible alone for interference, and that remaining silver ions also provide some interference. This is to be linked to the observation that zinc-stained proteins, although transparent, must be freed from zinc ion (so-called mobilization in [4]) before any further analysis by blotting or mass spectrometry can take place. Thus, heavy metal ions protein salts do not lead themselves easily to further analysis.

The conclusions of this work for future improvement is two-fold.
The first track for improvement lies in the availability of formaldehyde-free silver staining protocols for improved sequence coverage at high sensitivity and with all the simplicity of silver staining (especially in hardware) over fluorescent staining. Carbohydrazide was selected over other chemicals such as semicarbazide, thiosemicarbazide on the basis of more regular performances. Hydroxylamine also gave a positive staining, but it was much less sensitive. All of these chemicals did not lead to any peptide modification. However, we were not able with these chemicals to devise a high-contrast and wide scope silver staining protocol of the quality reached with formaldehyde. Decreasing the reducer concentration in the developer decreased the background but also dramatically decreased the sensitivity (data not shown). We also tried reducers used in photography. However, most of them are said to be "tanning", i.e. to induce strong protein crosslinking (e.g. hydroquinone, aminophenols, pyrogallol etc...), and were therefore not tested. Only a few molecules are known to be non-tanning, i.e. not to induce crosslinks. Among those are phenylene diamines and phenidone. However, we could not devise a practical silver staining protocol with any of the latter chemicals.

The second track for improvement is the development of improved destaining protocols. We now know that the main action of destaining protocols is not only the removal of the silver image but the oxidation of remaining formaldehyde and the complexation of silver ion. In the destaining protocols described to date, silver ion complexation is achieved by thiosulfate [8] or by ammonia [9], while silver and formaldehyde oxidation is achieved by hydrogen peroxide [9] or by ferricyanide [8], which is the active component of the classical Benedict's test solution for aldehydes.

Now that the chemical species responsible for interference have been identified, post-silver staining processing protocols with increased efficiency can probably be devised. They would allow even better compatibility between the high sensitivity and high contrast classical silver stain and the downstream peptide mass analysis.

On a more practical point of view, the formaldehyde-free silver staining described here combines the MS compatibility of colloidal Coomassie or fluorescent stains with the sensitivity of silver staining. It allows to reach the same performances in terms of identification than fluorescent probes, but with the permanence of the silver image and the simplicity of spot cutting from such an image. These are obvious practical advantages which combine to the much lower price of silver staining compared to commercial fluorescent probes (e.g. Sypro Ruby) to make this procedure an attractive choice. In addition, this procedure allows to perform a complete study with silver staining, i.e. both the analytical gels and the preparative work, thereby minimizing the confusing variations in spot intensities arising sometimes from changes in detection protocols. However, the poor staining homogeneity, when compared to either classical silver staining or colloidal Coomasie Blue or fluorescent probes, reduces the interest of this protocol for 2D gel-based proteomics. It could be successfully used for fast, sensitive and easy lane visualization in SDS-PAGE-MS/MS-based proteomics.

Table 1 : Comparison of the coverage efficiency by mass spectrometry after different detection methods

|  | Carbonic anhydrase | Ovalbumin | Bovine serum albumin | β Galactosidase | Glycogen phosphorylase |
|---|---|---|---|---|---|
| MALDI | Number of peptides/ %coverage | Number of peptides/ %coverage | Number of peptides/ %coverage | Number of peptides/ %coverage | Number of peptides/ %coverage |
| **Colloidal blue** | 6/ 32% | 12 /29% . | 16 /30% | 24 /27% | 36 /42% |
| **Sypro ruby** | 4 /23% | 8 /24% | 11 /19% | 20 /21% | 17 /19% |
| **Ruthenium** | 6 /32% | 6 /22% | 7 /12% | 10 /8% | 46 /47% |
| **Silver staining with HCHO, destained** | 8 /43% | Not identified | 5 /9% | 13 /10% | Not identified |
| **Silver staining with hydrazide, destained** | 7 /32% | 12 /35% | 17 /32% | 16 /20% | 17 /22% |
| **Silver staining with hydrazide, no destaining** | 6 /32% | 5 /16% | 8 /14% | 22 /23% | Not identified |
| LC/ESI-MS |  |  |  |  |  |
| **Colloidal blue** | 11 /38% | 9 /27% . | 33 /46% | 25 /27% | 32 /35% |
| **Sypro ruby** | 11 /29% | 5 /15% | 33 /47% | 14 /16% | 43 /48% |
| **Ruthenium** | 9 /26% | 8 /17% | 34 /57% | 17 /19% | 28 /28% |
| **Silver staining with HCHO, destained** | 6 /22% | 3 /5% | 24 /40% | N/A | 14 /14% |
| **Silver staining with hydrazide, destained** | 6 /20% | Not identified | 40 /55% | 20 /19% | 28 /34% |
| **Silver staining with hydrazide, no destaining** | 6 /20% | 4 /13% | 26 /42% | 12 /11% | 20 /23% |

The marker proteins described in the table are separated on a SDS gel, and detected by the method indicated. The 200 ng band is excised for each protein and then submitted to trypsin digestion. The peptide mixture is then analyzed either with MALDI/MS or with LC/ESI-MS, as described in the methods section.

Table 2 : determination of the useful threshold for various detection methods

|  | Carbonic anhydrase (number of peptides/sequence coverage) | | | | Glycogen phosphorylase (number of peptides/sequence coverage) | | | |
| --- | --- | --- | --- | --- | --- | --- | --- | --- |
|  | 200 ng | 100 ng | 50 ng | 20 ng | 200 ng | 100 ng | 50 ng | 20 ng |
| ruthenium | 5/18% | 4/15% | N.D. | N.D. | 18/23% | 16/22% | 14/19% | N.D. |
| F/FECN | 2/6% | N.I. | N.I. | N.I. | 16/28% | 12/15% | 2/2% | N.I. |
| F/perox | N.I. | N.I. | N.I. | N.I. | 13/21% | 4/5% | N.I. | N.I. |
| C/FECN | 4/14% | 4/14% | 2/6% | 2/6% | 18/21% | 13/15% | 7/11% | N.I. |

Serial dilutions of carbonic anhydrase and glycogen phosphorylase were separated by SDS PAGE and detected with various methods. The gel plugs were submitted to additional destaining if mentioned, and then analyzed by LC/ESI/MS of the peptide mixture resulting from in-gel trypsin digestion.
F: silver staining with formaldehyde developer, C: silver staining with carbohydrazide developer. FECN: ferricyanide-thiosulfate destaining process, Perox: hydrogen peroxide destaining process. Ruthenium: detection with the ruthenium-bathophenanthroline disulfonate complex.
N.D.: not detected on the gel; N.I: not identified (0 matching peptide on the LC/ESI/MS-MS spectrum)

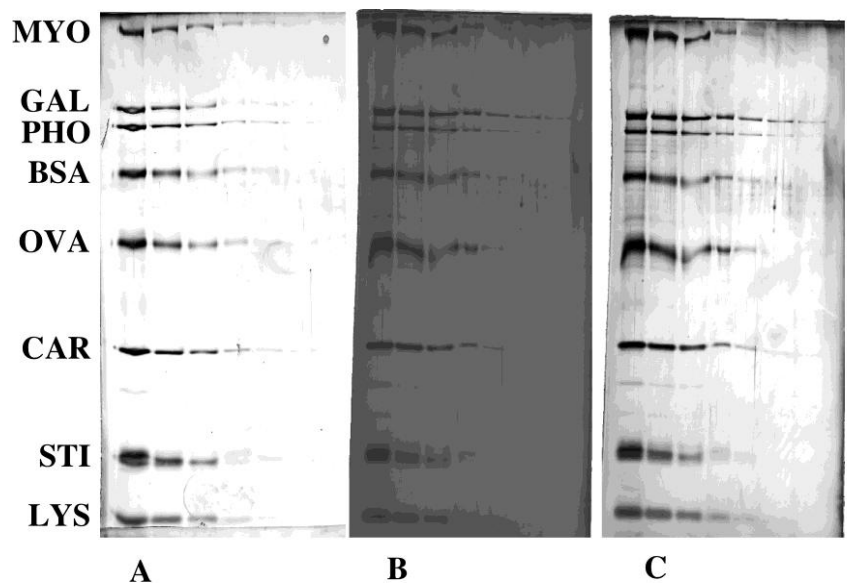

Figure.1. Sensitivity evaluation of the silver staining

Molecular weight markers (BioRad, broad range) were diluted serially in SDS buffer and separated by SDS electrophoresis. The corresponding gel was stained with standard (formaldehyde) silver staining (panel A), or with the carbohydrazide silver stain (panels B and C). The difference between panels B and C is a background substraction operated at the scanning stage. Protein separated: MYO: myosin (205kDa), GAL: beta galactosidase (116kDa), PHO: glycogen phosphorylase (97kDa) BSA: bovine serum albumin (67kDa), OVA: ovalbumin (46 kDa), CAR: carbonic anhydrase (30 kDa), STI: soybean trypsin inhibitor (21 kDa), LYS: lysozyme (14.5 kDa). Protein content per lane in each panel, from left to right: 200 ng/protein, 100 ng, 50 ng, 20 ng, 10 ng, 5 ng, 2 ng

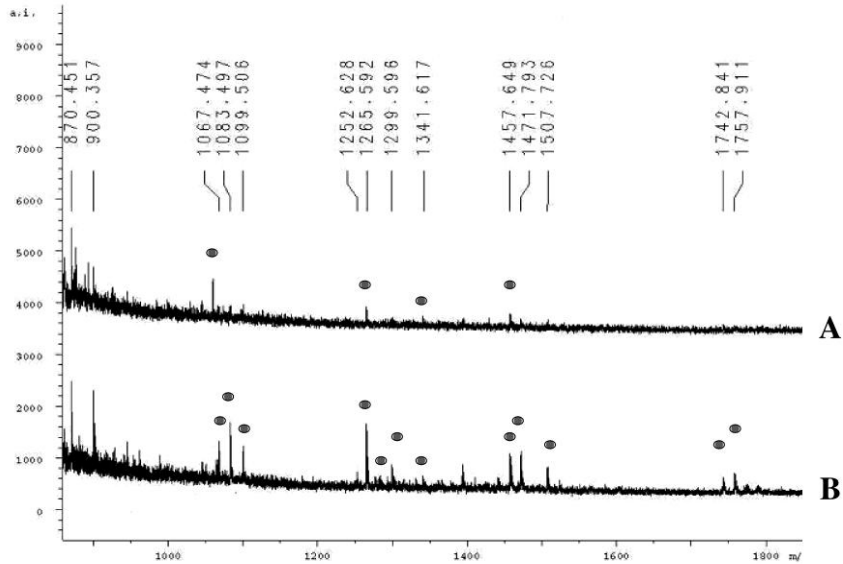

Figure. 2: Comparison of MALDI-MS Spectra after silver staining.
The Beta galactosidase band (ca. 2 picomoles) was excised from a silver-stained gel either stained with a formaldehyde developer and immediately destained with ferricyanide-thiosulfate (spectrum A) or stained with the carbohydrazide developer and not destained (spectrum B). The MALDI-MS spectra obtained after trypsin digestion of the corresponding bands are shown on the figure. The peaks matching with calculated masses for trypsin digestion of E. coli beta-galactosidase are marked with a solid circle

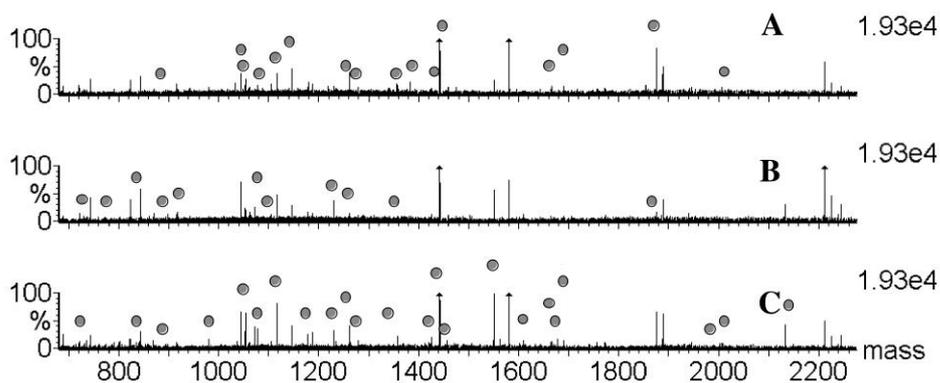

Figure. 3: Comparison of nanoLC ESI-MS Spectra after silver staining.
The glycogen phosphorylase band (2 picomoles) was excised from a silver-stained gel either stained with the carbohydrazide developer and immediately destained with ferricyanide-thiosulfate (spectrum A) or stained with a formaldehyde developer and destained (spectrum B). Spectrum C come from a ruthenium-stained gel. The combined nanoLC ESI-MS spectra (from retention time 5 to 40 min.) obtained on a Q-TOF II instrument (Micromass) after trypsin digestion of the corresponding bands are shown on the figure. Spectra are normalized in fonction of intensity. The peaks matching with calculated masses for trypsin digestion of glycogen phosphorylase are marked with a solid circle.

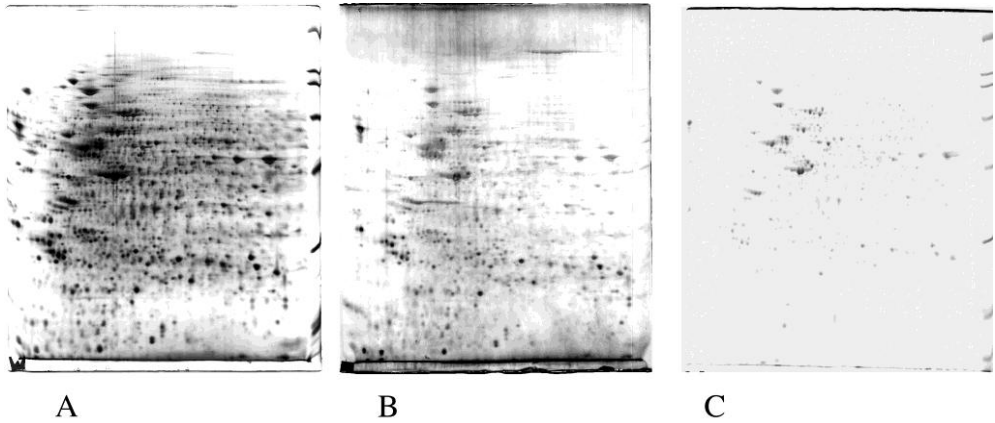

Figure 4: comparison of formaldehyde and carbohydrazide silver staining
A complex total cell extract from heLa cells was separated by 2D gel electrophoresis. 0.1 mg of proteins were loaded by in gel rehydration. IEF: immobilized pH gradient, pH 4 to 8, linear. SDS-PAGE: 10%T continuous gel.
A; detection with formaldehyde silver staining. B: detection with carbihydrazide silver staining. C: detection with colloidal Coomassie Blue